# Sustained electron self-injection in an evolving ellipsoid bubble for laser-plasma interaction

X. F. Li[1,2], Q. Kong[1]*, S. Kawata[2]*, Y. J. Gu[3], Q. Yu[1,3] and J. F. Qu[1]

[1] *Institute of Modern Physics, Fudan University, Shanghai 200433, People's Republic of China*

[2] *Graduate School of Engineering, Utsunomiya University, 7-1-2 Yohtoh, Utsunomiya 321-8585, Japan*

[3] *Institute of Physics of the ASCR, ELI-Beamlines Project, Na Slovance 2, 18221 Prague, Czech Republic*

## Abstract

Electron injection in an evolving ellipsoid bubble for laser wakefield acceleration is investigated by 2.5D PIC (Particle-In-Cell) simulation. Generally speaking, the self-injection electrons come from the position near the transverse radius in the bubble acceleration. However, we found the electrons near the laser axis also can be trapped into a longitudinal-expanding bubble. Moreover, this new self-injection mechanism is still existence after the electron self-injection stopped, which initially locates at near the bubble transverse radius. This phenomenon is confirmed through single-particle dynamic simulation. Besides, this new mechanism brings a high charge electron beam for acceleration, due to the sustained self-injection.

The laser wakefield acceleration (LWFA) has been studied many years since 1979, and it was first proposed by Tajima and Dowson[1]. With the development of laser technology, this research field has obtained abundant results[2]. In the early researches[3, 4], the electron beam has 100% energy spread. In 2004, the mono-energetic electron beams were firstly produced in the bubble regime by three groups[5-7]. Recently, electron beams were obtained in non-preformed plasma channels with an energy up to 2 GeV by using a 7cm-long gas cell[8]. An electron beam with energy of 4.2 GeV was reported by using 16 J of a laser and a plasma channel produced 9cm-long capillary discharge[9].

The electron bubble was firstly proposed by Pukhov *et al.* through PIC (Particle-In-Cell) simulation in 2002[10]. When an intense laser pulse penetrates through a low-density plasma, a bucket forms after all the plasma electrons interacting with the focused laser pulse are expelled by the radiation pressure, leaving the ions immobile. The charge-separation field attracts bulk electrons along the axis and forms a cavity devoid of electrons (bubble), which trails the driving laser beam at relativistic speed. Based on the research of W. Lu *et al.*, the stable bubble shape is sphere[11-13]. However, the general bubble is ellipsoid which depends on the laser and plasma parameters, and its shape also changes following laser propagated into plasma.[14-16].

If some plasma electrons were trapped into plasma wakefield or bubble regime, they can be accelerated and obtained high energy finally. The trapping process was named as the electron self-injection in laser plasma acceleration. In the previous researches, the majority of acceleration electrons come from the position near the transverse radius of bubble [17-19] and the charge of electron beam depends on the bubble geometry[14]. In this paper, it is found that some electrons from the position near the laser axis injected into the bubble and obtained high energy through PIC simulation. This phenomenon is different with the electron self-injection in the plasma wave, which was explained by 1D longitudinal wave breaking[20]. In our research, the election injection near the laser axis in bubble regime attributed to the evolving longitudinal radius of bubble and the laser field. This kind of electron injection is also confirmed through single particle dynamics.

The 2.5D PIC simulations were performed using the electromagnetic relativistic code "ZOHAR"[21]. The laser beam was linearly polarized in the $y$-direction with a Gaussian focused profile:

$$E = E_0 \frac{w_0}{w(x)} \exp(-\frac{y^2+z^2}{w^2(x)}) \exp(-\frac{(kx-\omega t)^2}{(0.5\tau)^2}) \cos(\varphi), \tag{1}$$

where $w(x) = w_0\left[1+(x-x_L)^2/z_R^2\right]^{0.5}$, $w_0$ is the laser waist, $x_L$ is the position of the laser waist, $z_R = \pi w_0^2/\lambda$ is the Rayleigh length, and $\tau = 21\,\text{fs}$ is the pulse duration. The laser wavelength was $\lambda = 800$ nm and the resolution was $dx = \lambda/32$ and $dy = \lambda_p/64$, where $\lambda_p$ is the plasma wavelength. The laser beam was focused at the left edge of the plasma. In the following analysis, the bubble radius is obtained by PIC simulation whereas the bubble velocity is set by theory to be $v_b = \sqrt{1-\gamma^{-2}}$ with $\gamma = 0.45\sqrt{n_c/n_0}$ [22].

When a laser ($a_0 = 5$ and $w_0 = 20\lambda$) interacted with an underdense plasma ($n_0 = 0.002n_c$), an electron bubble was formed. As shown in Fig. 1 (a) for simulation time at 600fs, it is a transversely ellipsoid bubble with the transverse radius is $R_\perp = 19.38\lambda$ and the longitudinal radius is $R_\parallel = 17.44\lambda$. Along with the laser propagating into plasma, laser profile was modulated by the plasma and it caused the bubble shape changing. At 2581fs, the transverse and longitudinal radiuses of bubble are 16.78 $\lambda$ and 25.00 $\lambda$, respectively. It is a longitudinally ellipsoid bubble, as presented in Fig. 1 (b). In this case, the bubble geometry changes from transversely ellipsoid to longitudinally ellipsoid following with time. Meanwhile, there are some acceleration electrons in the bubble and the energy spectrum of them at 2581fs was plotted in Fig. 1(c). The maximum of electron energy is 226 MeV and the energy spread is $\Delta E/E = 31.7\%$. Traced back to the initial position (denoted as $x_0, y_0$) of these energetic electrons ($E_i \geq 10 MeV$), as revealed in Fig.1 (d), one can find that there are some electrons came from the position near the transverse radius of bubble while many of electrons located near the laser axis initially. For simplicity, we named the first kind of electrons as far-axis injection electrons and the second kind of them as near-axis injection electrons. In this case, the far-axis injection electrons initially locate close to the plasma boundary while the near-axis injection electrons are distributing along the way of laser propagating, especially their initial position $x_0 > 500\lambda$.

There are three typical dynamics for the electrons around the bubble during the self-injection process, and their trajectories are analyzed through the PIC simulation as shown in Figs. 1(e) and (f). The first kind of them ($y_0$=40λ, denoted by the green line) is the near-axis injection electron and it can be trapped into the transversely ellipsoid bubble at 600fs. The second kind of electrons ($y_0$=30λ, denoted by the blue line) is initially around the transverse radius of bubble and is injected to the bubble. This injection process has been studied on the previous researches, and it was named as far-axis injection electron in this paper. The last kind of electrons ($y_0$=20λ, denoted by the black line) is pushed out by the laser field and forms the bubble electron sheath. It is noted that the motion of the near-axis injection electrons is always in the core of bubble, while the far-axis injection electrons forms the bubble electron sheath firstly and then injects into the bubble.

At 1651fs, the same three kinds of electrons are also analyzed in a longitudinally ellipsoid bubble, as presented in Fig. 1(f). The near-axis injection still occurred while the fax-axis injection has stopped. For the far-axis electron injection, based on the previous researches: the injection process could stop as laser propagating in the plasma[23, 24] and it also disappear when the bubble shape deviated seriously from a spherical shape[14]. In this case, when the far-axis self-injection stopped at 1651fs, it is sustainability for the near-axis injection. Then it is possible to increase the charge of acceleration electrons through the near-axis injection.

Based on the research of Kostyukov *et al.*[23], the bubble shape is assumed as a fixed sphere. When bubble radius and it velocity have the relationship $\gamma/R \leq 1/\sqrt{2}$, the electron near the bubble radius can be trapped into the bubble. Meanwhile, in the research of S. Kalmykov *et al.* [24], the bubble geometry is assumed as an expanding sphere. When the radius expands with enough large expansion rate, some electrons near the bubble radius can be injected into the bubble. After that, the bubble radius becomes stable or decreases, the self-injection stopped. In the both theories, it was assumed that the electrons near the laser axis were totally pushed out by laser field. However, when the laser intensity is weak, some electrons could pass though the laser field and move into the core of bubble[15]. It is possible for these electrons to be captured by the bubble and obtained high energy.

The single electron dynamic was used to confirm the near-axis self-injection and to analyze the condition for this phenomenon. According to the result of PIC simulation, as shown in Fig. 2(a), the bubble shape changes from transversely to longitudinally ellipsoid. The longitudinal radius of bubble is changing sharply compared with the transverse radius. For simplicity, it is assumed that the longitudinal radius of bubble is increasing with time while the transverse radius is stable ($R_\perp = 20.00\lambda$). In this case, the maximum value of growth rate for the longitudinal radius is $0.025\, \lambda/fs$.

Based on the theory of ellipsoid bubble[14], the electromagnetic field in bubble can be written as:

$$E_x = \left[(1-n_a) + v_b n_a v_a\right] \frac{\eta^2}{\eta^2 (1-v_b^2) + 2} \xi, \tag{2}$$

$$E_y = \frac{(2-\eta^2 v_b^2)(1-n_a) y}{2\eta^2 (1-v_b^2) + 4} - \frac{\eta^2 v_b n_a v_a}{2\eta^2 (1-v_b^2) + 4} y, \tag{3}$$

$$E_z = \frac{(2-\eta^2 v_b^2)(1-n_a) z}{2\eta^2 (1-v_b^2) + 4} - \frac{\eta^2 v_b n_a v_a}{2\eta^2 (1-v_b^2) + 4} z, \tag{4}$$

$$B_x = 0, \tag{5}$$

$$B_y = \frac{v_b \eta^2 (1-n_a) z}{2\eta^2 (1-v_b^2) + 4} + \frac{(2+\eta^2) n_a v_a}{2\eta^2 (1-v_b^2) + 4} z, \tag{6}$$

$$B_z = -\frac{v_b \eta^2 (1-n_a) y}{2\eta^2 (1-v_b^2)+4} - \frac{(2+\eta^2) n_a v_a}{2\eta^2 (1-v_b^2)+4} y, \qquad (7)$$

where, $\xi = x - v_b t$, $\eta = R_\perp / R_\parallel$. The bubble velocity is also set by theory to be $v_b = \sqrt{1-\gamma^{-2}}$ with $\gamma = 0.45\sqrt{n_c/n_0}$ [22]. The effect of residual electrons is ignored ($n_a = 0, v_a = 0$) in this paper. Here, we use dimensionless units, normalizing the length to $k_p$, the velocity to $c$, the electric field to $mc\omega_p/e$ and the magnetic field to $m\omega_p/e$.

The trajectory of an electron in the bubble was investigated by numerically solving the relativistic Newton-Lorentz equation, $\frac{d\mathbf{P}}{dt} = -e\left(\mathbf{E} + \frac{\mathbf{P}}{\gamma} \times \mathbf{B}\right)$, with the fourth-order Runge-Kutta method[23-27]. A schematic representation of an electron in the bubble field is presented in Fig. 2 (b). Considering the limited distribution of the bubble field, a modified factor $f(r) = \left[\tanh(R_\parallel/d - r/d) + 1\right]/2$ was multiplied with Eqs. (2)-(7) during calculation bubble field, where $r = \sqrt{\xi^2 + (y^2 + z^2)/\eta^2}$ and $d$ is the width of the electron sheath. In our calculation, $d = 0.5$ was used and the laser field was also introduced. The position of the core of laser pulsed located at the front of bubble. Initially, the electron at rest in front of the bubble with an initial position defined as $x_0$, $y_0$. To depict the process of the bubble evolving, the longitudinal radius is growing in time: $R_\parallel = R_\parallel^0 + \varepsilon \cdot t$ and the transverse radius is fixed: $R_\perp = 5.62$. The maximum value of the corresponding normalized growth rate for the case in Figs. 1 is $\varepsilon_{max} = 0.065$.

The trajectories of three electrons with different initial positions in the bubble coordinate system were presented in Fig. 2(c). The initially longitudinal radius of bubble is $R_\parallel^0 = 5.8$ and the growth rate is $\varepsilon = 0.06$, which is corresponding to the parameters in the Fig. 1(e). The electron with $y_0 = 5.87$ (blue line) was trapped in the bubble, which has been discussed in previous research[23]. At the same time, the electron with $y_0 = 2.89$ (green line) interacted with laser field firstly, entered into the bubble field and accelerated in the bubble finally. It is consistent with the result of PIC simulation as shown in Fig. 1(e). As presented in Fig. 2(d), the bubble shape is longitudinally ellipsoid initially with $R_\parallel^0 = 7.62$ and $\varepsilon = 0.06$. The electron with $y_0 = 5.62$ (blue line) was not trapped into the bubble, which means the far-axis electron injection has stopped. However, the electron located in $0 \leq y_0 \leq 0.39$ and $1.02 \leq y_0 \leq 1.39$ can be captured into the bubble. It means the near-axis electrons could inject into the bubble when the far-axis injection stopped.

For the near-axis electron injection, it should attribute to the bubble evolving and the laser field. As shown in Figs. 3(a) and (b), the near-axis self-injection did not appear in a fixed bubble ($\varepsilon = 0$). The other parameters are same as that of the initial parameters in Figs. 2(c) and (d). Besides, the near-axis self-injection also did not occur in an evolving sphere bubble without the laser field, as shown in Fig. 3(c). The bubble parameters are same as them in Ref. [24][24]. For the laser parameters, based on the research of W. Lu et al.[13], the laser intensity is $a_0 = R^2/4 = 6.25$, the laser waist is $w_0 = 50\lambda$ and the laser pulse is $\tau = 21 fs$, respectively. Including the laser field, it is found that the area of the far-axis injection was reduced from $3.20 \leq y_0 \leq 5.80$ to $4.21 \leq y_0 \leq 5.80$. At the same time, some electrons near the laser axis injected into the bubble, and the region of them is $0 \leq y_0 \leq 1.23$. Although the transverse radius of bubble is fixed in our assumption, the same phenomenon also appears.

As presented in Figs. 4, it was studied that the dependence of the region of the near-axis injection on the longitudinal radius growth rate (ε) in different shape bubbles. The relationship between the electron final energy and its initial position (y₀) was plotted in Fig. 4(a). The parameters are same as that of them in Fig. 2(c). The initial position of high energy electrons is the injection area for the bubble acceleration. For the far-axis injection, the minimum and the maximum of injection position were donated by $R_{far}^{min}$ and $R_{far}^{max}$, respectively. For the near-axis injection, $R_{near}^{min}$ and $R_{near}^{max}$ were used. In this case, $R_{near}^{min} = 2.89$, $R_{near}^{max} = 4.12$, $R_{far}^{min} = 5.87$ and $R_{far}^{max} = 6.39$, respectively. The far-axis injection area is from $R_{far}^{min}$ to $R_{far}^{max}$, as donated by red region in Fig. 4(b). With the growth rate of longitudinal radius increases, more electrons can be trapped into the bubble either through the fax-axis injection or the near-axis injection. When $\varepsilon \geq 0.14$, all of the electrons before the bubble could be injected into the bubble. It should own to the growth of longitudinal radius is faster than the electron motion in the bubble, where electron cannot escape from the bubble and it is impossible for the bubble acceleration. It is noted that the threshold value for the far-axis injection is $\varepsilon = 0.046$, while it is $\varepsilon = 0.05$ for the near-axis injection. For the transversely ellipsoid bubble ($R_{\parallel}^0 = 4.8$), as shown in Fig. 4(c), the near-axis electron injection appears when $\varepsilon \geq 0.07$. For the longitudinally ellipsoid bubble ($R_{\parallel}^0 = 7.6$), the threshold value is $\varepsilon = 0.04$, as revealed in Fig. 4(d). It means the near-axis electron injection appears facility in the longitudinal ellipsoid bubble.

At last, we compared the characteristics of acceleration electron beam in the case of near-axis injection with the case of far-axis electron injection. When a laser ($a_0 = 10$ and $w_0 = 5\lambda$) interacted with the plasma ($n_0 = 0.002 n_c$, the same plasma density as Figs. 1), the distribution of plasma density at 2581fs was plotted in Fig.5 (a). The initial position of acceleration electrons were plotted in Fig. 5(b) and the most of them distributed near the transverse radius. It means the near-axis injection did not happen in this case. Meanwhile, the energy spread of these acceleration electrons is 10.24% as presented in Fig. 5(c), which is much better than that of then in the case of Figs. 1. The number of acceleration electrons keeps stable after 1500fs for this case (green line), as shown in Fig. 5(d). While, the number of acceleration electrons increasing continuously following with time in the case of near-axis electron self-injection. It means the near-axis electron self-injection could produce a more electric quantity electron beam.

In this paper, we found that the electrons near the laser axis could be injected into the bubble and becomes acceleration electron. When the shape of bubble is unstable, this injection cross-section will appear, even the electron self-injection near the bubble transverse radius did not happen. It is easy found in the longitudinal ellipsoid bubble and this electron injection could produce a large electric quantity electron beam.

This work was partly supported by NSFC (No. 11175048), the Shanghai Nature Science Foundation (No. 11ZR1402700), and the Shanghai Scientific Research Innovation Key Projects No. 12ZZ011. The work was also partly supported by China Scholarship Council, Shanghai Leading Academic Discipline Project B107, the JSPS KAKENHI Grant Number 15K05359, MEXT, JSPS, the ASHULA project, ILE/Osaka University, CORE (Center for Optical Research and Education, Utsunomiya University, Japan), Fudan University, and CDI (Creative Dept. of Innovation, CCRD, Utsunomiya University).

# Reference

1. T. Tajima and J. M. Dawson, Phys. Rev. Lett. **43**, 267-270 (1979).
2. G. Mourou and D. Umstadter, Physics of Fluids B-Plasma Physics **4** (7), 2315-2325 (1992).
3. P. Mora, Physics of Fluids B-Plasma Physics **4** (6), 1630-1634 (1992).
4. J. B. Rosenzweig, B. Breizman, T. Katsouleas and J. J. Su, Phys. Rev. A **44** (10), R6189-R6192 (1991).
5. J. Faure, Y. Glinec, A. Pukhov, S. Kiselev, S. Gordienko, E. Lefebvre, J. P. Rousseau, F. Burgy and V. Malka, Nature **431** (7008), 541-544 (2004).
6. C. G. R. Geddes, C. Toth, J. van Tilborg, E. Esarey, C. B. Schroeder, D. Bruhwiler, C. Nieter, J. Cary and W. P. Leemans, Nature **431** (7008), 538-541 (2004).
7. S. P. D. Mangles, C. D. Murphy, Z. Najmudin, A. G. R. Thomas, J. L. Collier, A. E. Dangor, E. J. Divall, P. S. Foster, J. G. Gallacher, C. J. Hooker, D. A. Jaroszynski, A. J. Langley, W. B. Mori, P. A. Norreys, F. S. Tsung, R. Viskup, B. R. Walton and K. Krushelnick, Nature **431** (7008), 535-538 (2004).
8. X. Wang, R. Zgadzaj, N. Fazel, Z. Li, S. A. Yi, X. Zhang, W. Henderson, Y. Y. Chang, R. Korzekwa, H. E. Tsai, C. H. Pai, H. Quevedo, G. Dyer, E. Gaul, M. Martinez, A. C. Bernstein, T. Borger, M. Spinks, M. Donovan, V. Khudik, G. Shvets, T. Ditmire and M. C. Downer, Nat. Commun. **4** (2013).
9. W. P. Leemans, A. J. Gonsalves, H. S. Mao, K. Nakamura, C. Benedetti, C. B. Schroeder, C. Toth, J. Daniels, D. E. Mittelberger, S. S. Bulanov, J. L. Vay, C. G. R. Geddes and E. Esarey, Phys. Rev. Lett. **113** (24) (2014).
10. A. Pukhov and J. Meyer-ter-Vehn, Appl. Phys. B-Lasers Opt. **74** (4-5), 355-361 (2002).
11. W. Lu, C. Huang, M. Zhou, M. Tzoufras, F. S. Tsung, W. B. Mori and T. Katsouleas, Phys. Plasmas **13** (5), 056709 (2006).
12. W. Lu, C. Huang, M. Zhou, W. B. Mori and T. Katsouleas, Phys. Rev. Lett. **96** (16), 165002 (2006).
13. W. Lu, M. Tzoufras, C. Joshi, F. S. Tsung, W. B. Mori, J. Vieira, R. A. Fonseca and L. O. Silva, Phys. Rev. Spec. Top.-Accel. Beams **10** (6), 061301 (2007).
14. Y. J. G. X. F. Li, Q. Yu, S. Huang, F. Zhang, Q. Kong and S. Kawata, **21**, 073109 (2014).
15. H. C. Wu, B. S. Xie, S. Zhang, X. R. Hong, X. Y. Zhao and M. P. Liu, Phys. Plasmas **17** (11), 113103 (2010).
16. R. Sadighi-Bonabi and S. H. Rahmatollahpur, Phys. Plasmas **17** (3), 033105 (2010).
17. S. Y. Kalmykov, A. Beck, S. A. Yi, V. Khudik, B. A. Shadwick, E. Lefebvre and M. C. Downer, in *Advanced Accelerator Concepts*, edited by S. H. Gold and G. S. Nusinovich (Amer Inst Physics, Melville, 2010), Vol. 1299, pp. 174-179.
18. S. Y. Kalmykov, A. Beck, S. A. Yi, V. N. Khudik, M. C. Downer, E. Lefebvre, B. A. Shadwick and D. P. Umstadter, Phys. Plasmas **18** (5) (2011).
19. S. Y. Kalmykov, S. A. Yi, A. Beck, A. F. Lifschitz, X. Davoine, E. Lefebvre, V. Khudik, G. Shvets and M. C. Downer, Plasma Phys. Control. Fusion **53** (1) (2011).
20. E. Esarey, C. B. Schroeder and W. P. Leemans, Reviews of Modern Physics **81** (3), 1229-1285 (2009).
21. A. B. Langdon and B. F. Lasinski, Methods in Computational Physics:Advances in Research and Applications, edited by K. John (Elsevier,1976),Vol. 16, p. 327.
22. C. Benedetti, C. B. Schroeder, E. Esarey, F. Rossi and W. P. Leemans, Phys. Plasmas **20** (10), 103108 (2013).

23. I. Kostyukov, E. Nerush, A. Pukhov and V. Seredov, Phys. Rev. Lett. **103** (17) (2009).
24. S. Kalmykov, S. A. Yi, V. Khudik and G. Shvets, Phys. Rev. Lett. **103** (13) (2009).
25. I. Kostyukov, A. Pukhov and S. Kiselev, Phys. Plasmas **11** (11), 5256-5264 (2004).
26. S. A. Yi, V. Khudik, S. Y. Kalmykov and G. Shvets, Plasma Phys. Control. Fusion **53** (1) (2011).
27. J. Thomas, A. Pukhov and I. Y. Kostyukov, Laser Part. Beams **32** (02), 277-284 (2014).

**Figure Caption**

**Figure 1**

The distribution of electron density when a laser ($a_0 = 5$, $w_0 = 20\lambda$ and $\tau = 21fs$) interacted with an under-dense plasma with $n_0 = 0.002n_c$ at 600fs (a) and 2581fs(b). The energy spectrum (c) and the initial position (d) for acceleration electrons at 2581fs. The electron trajectories for different position electrons in different shape bubbles calculated using PIC simulation (e) t=600fs, (f) t=1651fs

**Figure 2**

The electron motion calculated through single particle dynamics. (a) The longitudinal radius and transverse radius of bubble as a function of time in the case of Figs. 1. (b) The geometry of single dynamics in the bubble field. The electron trajectories in a transverse bubble (c) and a longitudinally ellipsoid bubble (d). The parameters are corresponding to them in Fig. 1(e) and (f). The growth rate of longitudinal radius is $\varepsilon = 0.06$.

**Figure3**

The trajectories of electrons in the non-evolving transversely ellipsoid bubble (a) and longitudinally ellipsoid bubble (b). The parameters are same as them in Fig. 2(c) and(d), respectively. The trajectories of electrons in an evolving sphere bubble without laser field(c) and with laser field(d). The bubble parameters are same as them in Ref. [23]. The laser parameters are calculated based on the theory proposed by W. Lu *et al*. in Ref [13].

**Figure 4**

The electron trapping cross-area as a function of the growth rate of longitudinal radius in different shape bubble. (a) the relationship between the electrons final energy and their initial position. (b)The bubble parameters are same as them in Fig. 2(c). The trapping cross-area also calculated in the initially transverse ellipsoid(c) and the initially longitudinal ellipsoid bubble (d). The transverse radius is 5.8 and the initial longitudinal radius is 4.8 and 7.6, respectively.

**Figure 5**

Electron acceleration in a stable bubble where injection electrons majority came from the position near the transverse radius of bubble. (a)The distribution of electron density at 2581fs when a laser ($a_0 = 5$, $w_0 = 20\lambda$ and $\tau = 21 fs$) interacted with underdense plasma ($n_0 = 0.002 n_c$).(b)The distribution of initial position for the acceleration electrons. (c) The energy spectrum for the acceleration electrons in the stable bubble. (d)The electron number as a function of time for an evolving bubble (red line) and a stable bubble (green line).

**Figure 1**

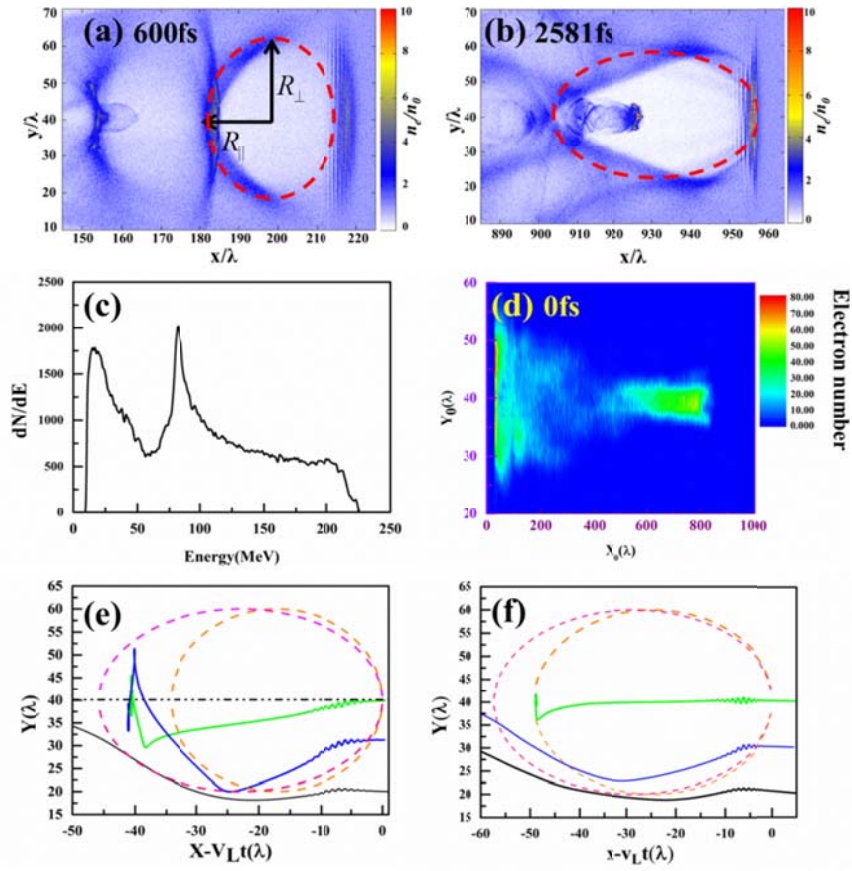

**Figure 2**

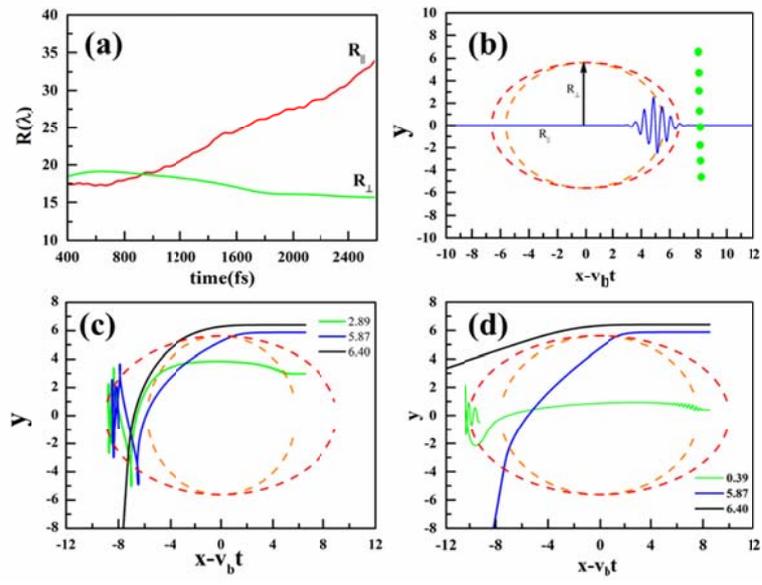

**Figure 3**

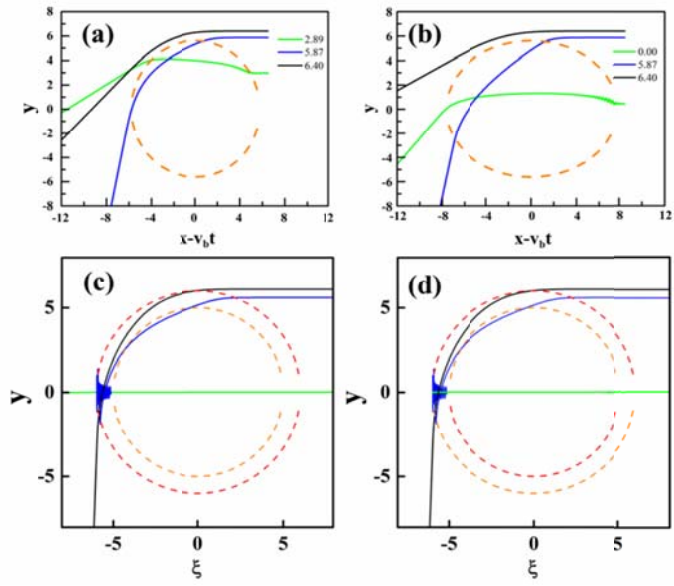

**Figure 4**

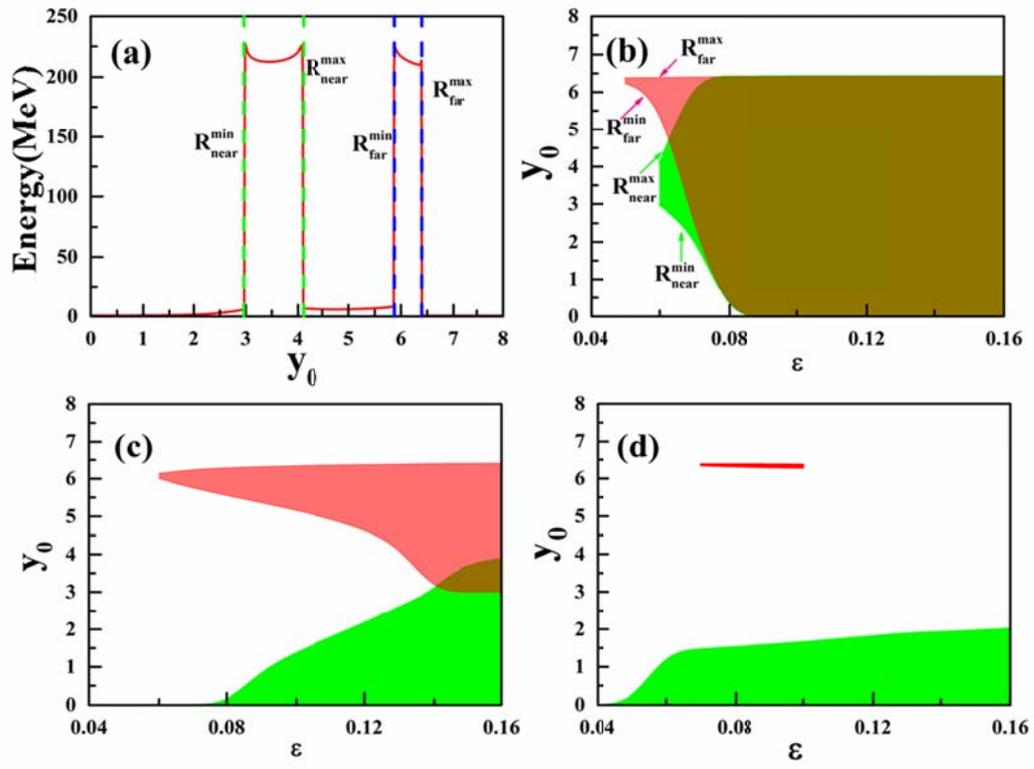

**Figure 5**

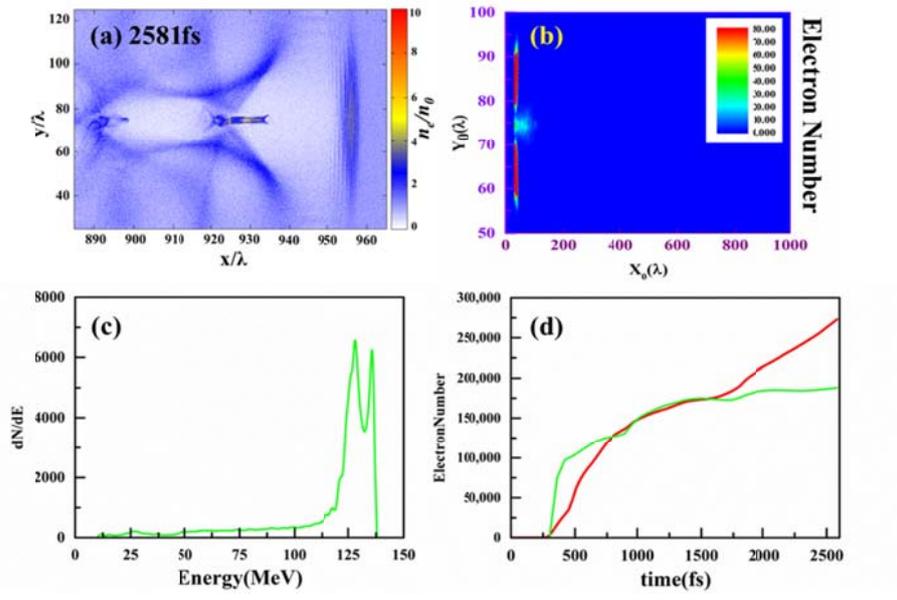